\documentclass[aps,preprint,superscriptaddress,12pt]{revtex4-2}

\usepackage{graphicx}
\usepackage{amsmath}
\usepackage{amssymb}
\usepackage{hyperref}
\usepackage{xcolor}
\usepackage[normalem]{ulem}

\begin{document}


\title{Force sensing with a graphene nanomechanical resonator coupled to photonic crystal guided resonances}

\author{Heng Lu}
\affiliation{School of Optoelectronic Science and Engineering \& Collaborative Innovation Center of Suzhou Nano Science and Technology, Soochow University, 215006 Suzhou, China}
\affiliation{Key Lab of Advanced Optical Manufacturing Technologies of Jiangsu Province \& Key Lab of Modern Optical Technologies of Education Ministry of China, Soochow University, 215006 Suzhou, China}

\author{Tingting Li}
\affiliation{School of Optoelectronic Science and Engineering \& Collaborative Innovation Center of Suzhou Nano Science and Technology, Soochow University, 215006 Suzhou, China}
\affiliation{Key Lab of Advanced Optical Manufacturing Technologies of Jiangsu Province \& Key Lab of Modern Optical Technologies of Education Ministry of China, Soochow University, 215006 Suzhou, China}

\author{Hui Hu}
\affiliation{Key Laboratory of Photonic Materials and Devices Physics for Oceanic Applications, Ministry of Industry and Information Technology, College of Physics and Optoelectronic Engineering, Harbin Engineering University, 150001 Harbin, China}
\affiliation{Key Laboratory of In-Fiber Integrated Optics of Ministry of Education, College of Physics and Optoelectronic Engineering, Harbin Engineering University, 150001 Harbin, China}

\author{Fengnan Chen}
\affiliation{School of Optoelectronic Science and Engineering \& Collaborative Innovation Center of Suzhou Nano Science and Technology, Soochow University, 215006 Suzhou, China}
\affiliation{Key Lab of Advanced Optical Manufacturing Technologies of Jiangsu Province \& Key Lab of Modern Optical Technologies of Education Ministry of China, Soochow University, 215006 Suzhou, China}

\author{Ti Sun}
\affiliation{School of Optoelectronic Science and Engineering \& Collaborative Innovation Center of Suzhou Nano Science and Technology, Soochow University, 215006 Suzhou, China}
\affiliation{Key Lab of Advanced Optical Manufacturing Technologies of Jiangsu Province \& Key Lab of Modern Optical Technologies of Education Ministry of China, Soochow University, 215006 Suzhou, China}

\author{Ying Yan}
\affiliation{School of Optoelectronic Science and Engineering \& Collaborative Innovation Center of Suzhou Nano Science and Technology, Soochow University, 215006 Suzhou, China}
\affiliation{Key Lab of Advanced Optical Manufacturing Technologies of Jiangsu Province \& Key Lab of Modern Optical Technologies of Education Ministry of China, Soochow University, 215006 Suzhou, China}

\author{Chinhua Wang}
\affiliation{School of Optoelectronic Science and Engineering \& Collaborative Innovation Center of Suzhou Nano Science and Technology, Soochow University, 215006 Suzhou, China}
\affiliation{Key Lab of Advanced Optical Manufacturing Technologies of Jiangsu Province \& Key Lab of Modern Optical Technologies of Education Ministry of China, Soochow University, 215006 Suzhou, China}
 
\author{Joel Moser}
\email{j.moser@suda.edu.cn}
\affiliation{School of Optoelectronic Science and Engineering \& Collaborative Innovation Center of Suzhou Nano Science and Technology, Soochow University, 215006 Suzhou, China}
\affiliation{Key Lab of Advanced Optical Manufacturing Technologies of Jiangsu Province \& Key Lab of Modern Optical Technologies of Education Ministry of China, Soochow University, 215006 Suzhou, China}

\begin{abstract} 
Achieving optimal force sensitivity with nanomechanical resonators requires the ability to resolve their thermal vibrations. In two-dimensional resonators, this can be done by measuring the energy they absorb while vibrating in an optical standing wave formed between a light source and a mirror. However, the responsivity of this method -- the change in optical energy per unit displacement of the resonator -- is modest, fundamentally limited by the physics of propagating plane waves. We present simulations showing that replacing the mirror with a photonic crystal supporting guided resonances increases the responsivity of graphene resonators by an order of magnitude. The steep optical energy gradients enable efficient transduction of flexural vibrations using low optical power, thereby reducing heating. Furthermore, the presence of two guided resonances at different wavelengths allows thermal vibrations to be resolved with a high signal-to-noise ratio across a wide range of membrane positions in free space. Our approach provides a simple optical method for implementing ultrasensitive force detection using a graphene nanomechanical resonator.
\end{abstract}

\maketitle
\section{Introduction}

Nanomechanical resonators based on suspended membranes of two-dimensional (2D) materials can be used to detect a wide range of external stimuli with high sensitivity \cite{Lemme2020,Steeneken2021,Xu2022,Ferrari2023}. One such application is the detection of nearly resonant forces \cite{Weber2016}. These 2D membranes are well suited for measuring weak forces because their low mass $m$ results in a low damping rate $\gamma=m\omega_0/Q$, where $\omega_0$ is the angular resonant frequency and $Q$ is the quality factor of the fundamental vibration mode. In turn, the power spectral density (PSD) of thermal forces driving the mode is small: $S=4k_\mathrm{B}T\gamma$, where $k_\mathrm{B}$ is the Boltzmann constant and $T$ is the modal temperature. This low thermal force background enables detection of nonthermal forces through their effects on the mechanical response.

However, achieving ultrasensitive force sensing requires more than just a resonator with extremely small mass; it also requires the ability to detect vanishingly small changes in the mechanical response \cite{Rugar2004,Degen2009,Moser2013,Eichler2022}. Various techniques have been developed for this purpose, including electrical transport \cite{Nathanson1967,Sazonova2004,Chen2009}, atomic force microscopy \cite{Garcia2008}, capacitive measurements \cite{Hertzberg2010,Song2012,Weber2014,Singh2014}, and optical methods \cite{Carr1997,Hadjar1999,Karabacak2005,Bunch2007,Reserbat2012,Stapfner2013,Cole2015,Schwarz2016,Barnard2019,Kirchhof2022}. Among these, the most sensitive vibrational readout is achieved using approaches that either couple the resonator’s vibrations to the electromagnetic mode of a cavity \cite{Weber2014,Singh2014,Stapfner2013,Cole2015,Barnard2019} or embed the resonator in a Michelson interferometer \cite{Schwarz2016,Kirchhof2022}. These methods benefit from a high responsivity and a low instrument noise background, allowing the use of weak probe signals that minimally perturb the vibrations.

While cavity optomechanics and interferometry are techniques of choice for ultrasensitive force detection, their implementation can be technically demanding. If higher instrument noise is acceptable, standard optical reflectometry \cite{Bunch2007} offers a simpler alternative that still provides sufficient sensitivity to detect thermal forces, especially near room temperature \cite{Lee2018}. This method involves placing the resonator in an optical standing wave, formed between a mirror and the light source, and measuring modulations in the reflected light intensity using a photodetector. Even though the reflectance of a membrane as thin as graphene is low, the reflectance $R$ of a device composed of a membrane, a mirror, and the vacuum gap between them can be high. $R$ depends on the membrane's position within the standing wave, that is, on the distance $z_\mathrm{m}$ between the membrane and the mirror. The dependence of $R$ on $z_\mathrm{m}$ arises from the spatial distribution of optical energy in the standing wave. For a perfectly reflective mirror, this distribution causes the absorbance $A\simeq1-R$ of the membrane to vary with $z_\mathrm{m}$ (Refs.~\cite{Favero2008,Chen2018}). As the membrane vibrates within the standing wave, changes in its position lead to modulations in the absorbed optical energy, which in turn modulate the reflected light intensity. Therefore, placing the resonator in a region with a steeper gradient of optical energy enhances the transduction of weaker vibrations.

In this work, we demonstrate through simulations that the responsivity $\mathcal{R}=\vert\partial R/\partial z_\mathrm{m}\vert$ of a graphene nanomechanical resonator is significantly enhanced when it is placed in the near field of a photonic crystal (PC) guided resonance, compared to when it is suspended over a planar mirror. Unlike experimental setups based on cavity optomechanics \cite{Aspelmeyer2014}, where vibrations are detected by monitoring the spectral response of a cavity, we consider a configuration in which monochromatic light is normally incident on the resonator suspended above the PC, and the intensity of the reflected light is measured. In essence, we replace the mirror in a standard reflectometry setup with a photonic crystal. First, we simulate the reflectance spectrum of the PC without graphene and identify two guided resonances -- that is, two guided modes \cite{Johnson1999} coupled to external radiation \cite{Peng1996,Kanskar1997,Astratov1999,Cowan2001,Fan2002} -- at distinct wavelengths. Next, we simulate the reflectance $R$ of the device composed of the graphene membrane, the PC, and the vacuum gap between them, as well as the absorbance of graphene, as functions of $z_\mathrm{m}$ for both wavelengths. At values of $z_\mathrm{m}$ where $\mathcal{R}$ is maximized, we calculate the spectrum of vibrational amplitude fluctuations due to thermal vibrations of graphene at room temperature. Taking into account the instrument noise of a commercial photodetector, we find that the signal-to-noise ratio (SNR) of the thermal spectrum is significantly higher when using the PC compared to a planar mirror. Because the maximum of $\mathcal{R}$ occurs at different $z_\mathrm{m}$ values for the two guided resonances, the SNR remains high even as $z_\mathrm{m}$ is varied by a large amount using a gate voltage. Overall, we demonstrate that replacing a planar mirror with a photonic crystal in a reflectance measurement enables a simple optical method for implementing ultrasensitive force detection with a graphene nanomechanical resonator. 

\section{Working principle and optical response of the photonic crystal}

We begin with a brief overview of the principle behind detecting the vibrations of 2D resonators using reflectometry. The output of the photodetector is a voltage $\delta V$ whose dependence on time $t$ reads     
\begin{equation}
\delta V(t)=\mathcal{T}P_\mathrm{inc}\left[R\vert_{z_\mathrm{m}}+\frac{\partial R}{\partial z}\Big\vert_{z_\mathrm{m}}\delta z(t)\right]\ast h(t) + \delta u(t)\,,\label{deltaV}
\end{equation}
where $\mathcal{T}$ is the transmittance of the optical path between the resonator and the photodetector, $P_\mathrm{inc}$ is the optical power incident on the resonator, $R$ is the reflectance of the device composed of the resonator, the mirror, and the vacuum gap between them, $z$ is the distance measured along the optical axis from the mirror, $\delta z$ is the vibrational amplitude, and $h$ is the response of the photodetector. The term $\delta u$ represents instrument noise. To ensure that $\delta z(t)$ is transduced into a measurable signal $\delta V(t)$, the responsivity $\mathcal{R}=\vert\partial R/\partial z\vert_{z_\mathrm{m}}$ must be large. In the case of a perfectly reflective mirror, the largest responsivity for a single-layer graphene resonator depends on the wavelength of the light $\lambda$ and reads
\begin{equation}
\mathcal{R}_\mathrm{max}\simeq\frac{8\pi}{\lambda}\times\frac{\sigma\eta}{1+2\sigma\eta}\,,\label{Rmaxmirror}
\end{equation}
where $\sigma$ is the intrinsic optical conductivity of graphene in the visible range and $\eta$ is the wave impedance of free space. A shorter $\lambda$ yields a larger $\mathcal{R}_\mathrm{max}$, although experimental constraints make the enhancement modest. Alternatively, the measurement described by Eq.~(\ref{deltaV}) can be interpreted as being enabled by the absorbance $A=1-R$ of the resonator. Indeed, the modulation of $\delta V(t)$ is primarily caused by changes in the optical energy absorbed by the membrane as it vibrates in the standing wave. The responsivity can then be rewritten as:
\begin{equation}
\mathcal{R}=\Big\vert\frac{\partial A}{\partial z}\Big\vert_{z_\mathrm{m}}\simeq\frac{\sigma\eta}{\vert\mathbf{E}_\mathrm{inc}\vert^2}\Big\vert\frac{\partial\vert\mathbf{E}\vert^2}{\partial z}\Big\vert_{z_\mathrm{m}}\,, \label{responsivity}
\end{equation}
where $\mathbf{E}_\mathrm{inc}$ is the electric field intensity of the incident light and $\mathbf{E}$ is that of the standing wave in the absence of graphene \cite{Mao2019}. The single-pass absorbance of graphene in free space is $\sigma\eta\simeq0.023$. Equation~(\ref{responsivity}) suggests that $\mathcal{R}$ can be enhanced simply by placing graphene in a strong gradient of optical energy.

\begin{figure}[t]
\centering\includegraphics{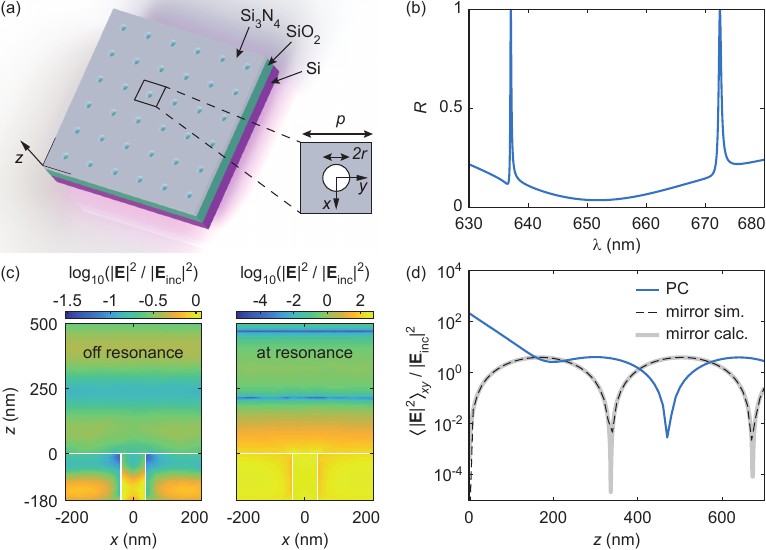}
\caption{Optical response of the photonic crystal. (a) Schematic of the photonic crystal (PC) and its substrate. The PC consists of a square lattice of holes patterned in a silicon nitride slab. Unit cell periodicity $p=400$~nm; hole radius $r=40$~nm; slab thickness $d=180$~nm. The slab is grown on oxidized silicon with an oxide layer thickness of 1200~nm. (b) Reflection spectrum $R(\lambda)$ of the PC for normally incident light with wavelength $\lambda$. (c) Spatial distribution of the squared magnitude of the electric field intensity $\vert\mathbf{E}\vert^2$, normalized to that of the incident field $\vert\mathbf{E}_\mathrm{inc}\vert^2$, in the $(xz)$ plane across the center of a hole for $\lambda=660$~nm (off resonance) and $\lambda=672.5$~nm (at resonance). White lines indicate the PC surface facing free space ($z=0$) and the cross-section of a hole. (d) $\langle\vert\mathbf{E}\vert^2\rangle_{xy}(z)/\vert\mathbf{E}_\mathrm{inc}\vert^2$, where $\vert\mathbf{E}\vert^2(z)$ is averaged over each $(xy)$ plane within a single unit cell. Simulations performed with the PC structure (``PC'') and with a perfectly reflective mirror (``mirror sim.'') are shown using $\lambda=672.5$~nm, alongside an analytical calculation for the mirror (``mirror calc.'') used to validate the simulations.}
\end{figure}

Large optical energy gradients can be found near the surface of PCs when their guided resonances are excited by incident light. Guided resonances are optical modes that, while confined within the crystal, are coupled to free-space radiation, causing some of their energy to leak out. We simulate such guided resonances using the structure depicted in Fig.~1a. We use CST Studio Suite, a three-dimensional electromagnetic field simulation software. The PC is a square lattice of holes in a silicon nitride slab. The slab is grown on an oxidized silicon substrate. We choose a lattice periodicy $p=400$~nm, a hole radius $r=40$~nm, and a slab thickness $d=180$~nm. Increasing $p$ shifts guided resonances to longer wavelengths. Increasing $r$ or decreasing $d$ broadens the linewidth of resonances in the reflectance spectrum, indicating larger energy leakage rates into free space. Light is normally incident from the top. Figure~1b displays the reflectance $R$ of the PC as a function of incident light wavelength $\lambda$. We find that $R(\lambda)$ exhibits a slowly modulated background with two sharp, asymmetric peaks superimposed on it near 637~nm and 672~nm. As in previous works \cite{Fan2002}, we attribute the background to Fabry-P\'{e}rot resonances in the structure, away from the guided resonances. We have verified that it matches $R(\lambda)$ for an unpatterned silicon nitride slab. The peaks, which reach $R\simeq1$, originate from guided resonances. Their asymmetry arises from interferences between light that is directly reflected by the slab (which accounts for the background) and light that first excites the guided resonances and then leaks out \cite{Fan2002}. Figure~1c shows the spatial distribution of the squared magnitude of the electric field intensity $\vert\mathbf{E}\vert^2$, normalized to that of the incident field $\vert\mathbf{E}_\mathrm{inc}\vert^2$, both within the PC and in the free space above it. Both the off-resonance ($\lambda=660$~nm) and the resonant cases ($\lambda=672.5$~nm) are shown. At resonance, $\vert\mathbf{E}\vert^2/\vert\mathbf{E}_\mathrm{inc}\vert^2$ is large in a region extending $\simeq100$~nm above the PC and decays with distance $z$ into free space. To better visualize this decay, in Fig.~1d we show $\langle\vert\mathbf{E}\vert^2\rangle_{xy}(z)/\vert\mathbf{E}_\mathrm{inc}\vert^2$, where $\vert\mathbf{E}\vert^2(z)$ is averaged over each $(xy)$ plane. This ratio can be interpreted as the optical energy of the standing wave normalized by the energy of the incident light, $U/U_\mathrm{inc}$. For comparison, the same quantity calculated for the case where the PC is replaced with a perfectly reflective mirror is also shown. In the far field, both the PC and the mirror display a similar trend in $U/U_\mathrm{inc}(z)$, albeit with a shift in $z$, and reach the expected maximum value of $U/U_\mathrm{inc}(z)\simeq4$ (Ref.~\cite{Chen2018}). In the near field, however, the PC and the mirror exhibit opposite behavior: the PC shows a large energy due to leakage from its guided resonances, while the mirror shows a low energy resulting from the node of the standing wave at its surface.

\section{Coupling of graphene to guided resonances and device responsivity}

Having characterized the response of the bare photonic crystal, we now turn to the device consisting of a graphene membrane suspended above it, as shown in Fig.~2a. A spacer (here, a patterned layer of boron nitride) is used to set the distance $z_\mathrm{m}$ between the top surface of the PC (defined as $z=0$ in Fig.~1a) and the membrane. Graphene, in contact with an electrode, and the doped silicon substrate together form a parallel-plate capacitor, enabling electrostatic tuning of $z_\mathrm{m}$ later in this work. 

Figure 2b shows that the reflectance spectra $R(\lambda)$ of both guided resonances are strongly affected by the presence of graphene. The reflectance $R$ is simulated at the surface of graphene facing the light source. As $z_\mathrm{m}$ decreases, $R$ is suppressed, as graphene couples to the guided resonances and acts as a dissipation channel for the PC. The reflectance approaches zero near $z_\mathrm{m}\simeq0$, a behavior consistent with the condition for critical coupling reported in Ref.~\cite{Piper2014}, while the peak wavelengths shift by only a fraction of a nanometer.

The presence of graphene also affects the spatial distribution of $\langle\vert\mathbf{E}\vert^2\rangle_{xy}/\vert\mathbf{E}_\mathrm{inc}\vert^2$ in free space, which exhibits weaker modulations as $z_\mathrm{m}$ decreases (Fig.~2c). In comparison, the distribution calculated for the planar mirror remains unchanged. As a consequence of the large near-field energy of the PC, the absorbance $A$ of graphene shows a pronounced dependence on $z_\mathrm{m}$ (Fig.~2d). At its maximum, $A$ is nearly 7 times larger than with the planar mirror. We have verified that $A$ is directly related to the normalized optical energy of the standing wave shown in Fig. 2c, namely
\begin{equation}
A\simeq\sigma\eta\frac{\langle\vert\mathbf{E}\vert^2\rangle_{xy}}{\vert\mathbf{E}_\mathrm{inc}\vert^2}=\sigma\eta\frac{U}{U_\mathrm{inc}}\,.\label{absorbance}
\end{equation}

The reflectance of the device also exhibits a strong dependence on $z_\mathrm{m}$ (Fig.~2e). The suppression of $R$ near $z_\mathrm{m}=0$ is consistent with the behavior shown in Fig.~2b. The steep decay of $R$ as $z_\mathrm{m}$ decreases translates into a high responsivity $\mathcal{R}(z_\mathrm{m})$, which is nearly one order of magnitude higher than with the mirror (Fig.~2f). We can now estimate and compare the force-sensing performance of devices in which graphene is suspended over either the PC or the mirror.

\begin{figure}[h]
\centering\includegraphics{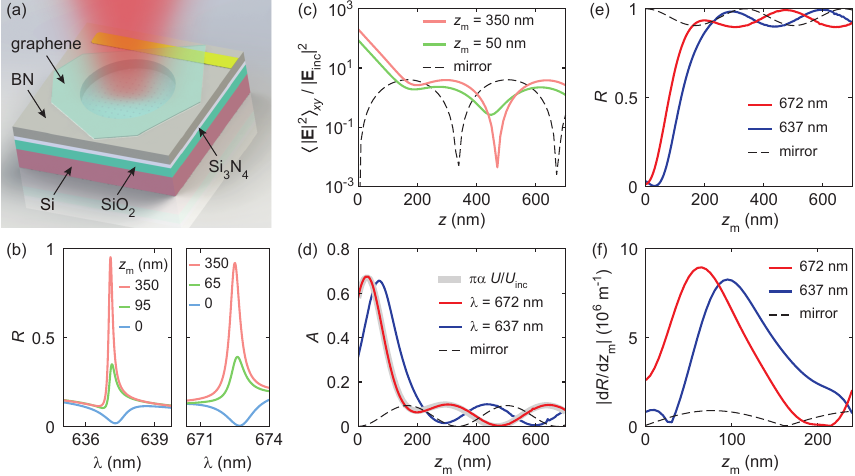}
\caption{Coupling of graphene to guided resonances and device responsivity. (a) Device schematic showing graphene suspended above a photonic crystal (PC). BN (boron nitride) acts as a spacer between graphene and the PC. Graphene contacts an electrode (yellow structure). (b) Reflectance spectra $R(\lambda)$ of the two guided resonances for three different values of $z_\mathrm{m}$, the distance between graphene and the PC. (c) Spatial distribution of $\langle\vert\mathbf{E}\vert^2\rangle_{xy}/\vert\mathbf{E}_\mathrm{inc}\vert^2$ in free space for the PC (solid lines) and the mirror (dashed line, $\lambda=672$~nm). The distribution for the mirror is independent of $z_\mathrm{m}$. (d) Absorbance $A$ of graphene as a function of $z_\mathrm{m}$ for the PC at the two peak wavelengths shown in (b) (blue and red lines), and for the mirror at $\lambda=672$~nm (dashed line). Also shown is $\sigma\eta U/U_\mathrm{inc}$, defined in the main text, as a function of $z_\mathrm{m}$ at $\lambda=672$~nm (thick, gray trace). Note that, for the PC, $A(z_\mathrm{m})$ and $\sigma\eta U/U_\mathrm{inc}(z_\mathrm{m})$ exhibit a maximum despite the absence of a maximum in $\langle\vert\mathbf{E}\vert^2\rangle_{xy}(z)$ at any fixed $z_\mathrm{m}$ [see panel (c)]. (e) Reflectance $R(z_\mathrm{m})$ of the device based on the PC (solid lines) and on the mirror (dashed line, $\lambda=672$~nm). (f) Responsivity $\mathcal{R}=\vert\partial R/\partial z_\mathrm{m}\vert(z_\mathrm{m})$ of the device based on the PC (solid lines) and on the mirror (dashed line, $\lambda=672$~nm).}
\end{figure}

\section{Sensing of thermomechanical force noise}

The smallest forces that a nanomechanical resonator can measurably respond to are Langevin forces responsible for the Brownian motion of its flexural vibrations in a thermal bath. Thermal vibrations of nanomechanical resonators based on 2D membranes have been measured optically, most commonly at room temperature and, in some cases, at cryogenic temperatures. Table~I in Appendix presents a survey of experimental studies in which such thermal vibrations have been measured. Here, we simulate an experiment in which the PSD of the thermal amplitude fluctuations of a graphene resonator, $S_{zz}$, is measured using reflectometry. We consider devices based on both the PC and the mirror configuration, and we estimate the signal-to-noise ratio (SNR) of measurements performed with a commercial photodetector (APD130A2 by Thorlabs). The single-sided PSD of voltage fluctuations $\delta u$ of nonmechanical origin [see Eq.~(\ref{deltaV})] at the detector's output is denoted by $S_{uu}$. We have measured $S_{uu}$ in an actual experimental setup and found that it increases linearly with $P_\mathrm{inc}$ [Ref.~\cite{Lu2021}] as:
\begin{equation}
S_{uu}\simeq a\times R\times\mathcal{T}\times P_\mathrm{inc}+b\,,\label{Suu}
\end{equation}
with $\mathcal{T}\simeq0.5$, $a\simeq6.5\times10^{-7}$~V$^2$~Hz$^{-1}$~W$^{-1}$, and $b\simeq3.5\times10^{-14}$~V$^2$~Hz$^{-1}$. The expression~(\ref{Suu}) is valid until the photodetector saturates near $R\times\mathcal{T}\times P_\mathrm{inc}\simeq2\times10^{-6}$~W. We therefore set $P_\mathrm{inc}=2\times10^{-6}$~W for both the PC-based and mirror-based devices. We define the SNR as:
\begin{equation}
\mathrm{SNR}=\frac{S_{vv}(f_0)}{S_{uu}}\,,\label{SNR}
\end{equation}
where $f_0=\omega_0/2\pi$ is the resonant frequency of the fundamental vibrational mode, and $S_{vv}=\beta^2S_{zz}+S_{uu}$ is the single-sided PSD of voltage fluctuations at the output of the photodetector. Here, $\beta=P_\mathrm{inc}\times\mathcal{T}\times\mathrm{G}\times\mathcal{R}$ is the transduction factor (in V~m$^{-1}$), with $\mathrm{G}\simeq1.25\times10^6$~V~W$^{-1}$ the transimpedance gain of the photodetector. According to the fluctuation-dissipation theorem, the PSD of vibrational amplitude fluctuations at resonance is $S_{zz}(f_0)=k_\mathrm{B}TQ/(2\pi^3mf_0^3)$. For a drum-shaped resonator under a radial stretching force per unit length $\xi$, the resonant frequency scales as $f_0\propto[\xi/(R_\mathrm{drum}^2\rho)]^{1/2}$, with $R_\mathrm{drum}$ the radius of the drum in m and $\rho$ its 2D mass density in kg~m$^{-2}$ (Ref.~\cite{Weber2016}). Combining these results, a high SNR requires not only a high $Q$ and a low instrument noise $S_{uu}$, but also a large ratio of $(m/\xi^3)^{1/2}$. To meet this condition with realistic parameters, we choose $R_\mathrm{drum}=4$~$\mu$m and $\xi=0.03$~N~m$^{-1}$, consistent with the single-layer graphene device measured in Ref.~\cite{Nicholl2015}.

\begin{figure}[t]
\centering\includegraphics{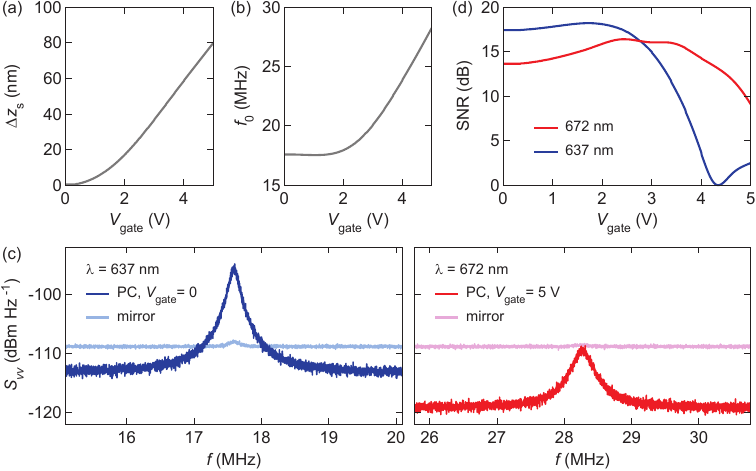}
\caption{Sensing of thermomechanical force noise at room temperature. (a) Calculated static displacement $\Delta z_\mathrm{s}=100-z_\mathrm{m}$ (nm) of the graphene membrane as a function of gate voltage $V_\mathrm{gate}$ for the structure depicted in Fig.~2a. (b) Calculated resonant frequency of the fundamental mode $f_0(V_\mathrm{gate})$. In (a, b), the graphene drum has the following properties: radius $R_\mathrm{drum}=4$~$\mu$m; radial stretching force per unit length $\xi=0.03$~N~m$^{-1}$; Young modulus $E=50$~N~m$^{-1}$; and Poisson ratio $\nu=0.165$. (c) Calculated thermal response $S_{vv}=\beta^2S_{zz}+S_{uu}$ as a function of spectral frequency $f$ for both the PC-based and the mirror-based devices. Results are shown for $\lambda=637$~nm at $V_\mathrm{gate}=0$ ($f_0\simeq17.6$~MHz, left-hand-side panel), and for $\lambda=672$~nm at $V_\mathrm{gate}=5$~V ($f_0\simeq28.3$~MHz, right-hand-side panel). Background levels vary with absorbance. Power of light incident on the resonator: $P_\mathrm{inc}=2$~$\mu$W. For both panels, the resolution bandwidth is 1.53~kHz. Averages are taken over 100 realizations of $S_{vv}(f)$ for the PC-based device and 1000 realizations for the mirror-based device. (d) Signal-to-noise ratio $\mathrm{SNR}(V_\mathrm{gate})$, for both wavelengths, calculated for the PC-based device according to Eq.~(\ref{SNR}).}
\end{figure}

Considering the geometry of Fig.~2a, we assume that a bias voltage $V_\mathrm{gate}$ is applied between graphene and the silicon acting as a gate electrode, resulting in a static displacement at the center of the membrane $\Delta z_\mathrm{s}(V_\mathrm{gate})$ [Fig.~3a] and in a tuning of the resonant frequency $f_0(V_\mathrm{gate})$ [Fig.~3b]. We assume that, at $V_\mathrm{gate}=0$, the membrane is flat and $z_\mathrm{m}=100$~nm, so that $\Delta z_\mathrm{s}=100-z_\mathrm{m}$ (in nm). We obtain $\Delta z_\mathrm{s}$ and $f_0$ by minimizing the total energy -- that is, the sum of the stretching elastic energy of the membrane and the capacitive energy stored in the capacitor defined by graphene, the silicon substrate, and the dielectric materials between them. In this calculation, we use a Young's modulus for graphene $E=50$~N~m$^{-1}$, as estimated in Ref.~\cite{Nicholl2015}, and a Poisson ratio $\nu=0.165$. Estimating $\Delta z_\mathrm{s}$ serves two purposes. First, it enables optimization of the device's responsivity. Second, it provides a means to quantify the change in vibrational temperature resulting from the coupling between vibrations and photothermal forces in the presence of a large optical energy gradient \cite{Metzger2004,Zaitsev2011}, such as that found in a bent membrane \cite{Barton2012}. These considerations apply only to the PC-based device. By contrast, we assume that the mirror-based device is always optimized to achieve the maximum responsivity, as defined by Eq.~(\ref{Rmaxmirror}). This assumption enables a direct comparison between the performance of the PC-based device and the optimal performance of the mirror-based device.

Figure~3c shows $S_{vv}$ as a function of the spectral frequency $f$. We assume that the resonator is at room temperature and has a quality factor $Q=100$, a value commonly measured in graphene resonators at this temperature. The left-hand-side panel corresponds to $\lambda=637$~nm at $V_\mathrm{gate}=0$, and the right-hand-side panel to $\lambda=672$~nm at $V_\mathrm{gate}=5$~V. At both wavelengths, dark-colored traces represent $S_{vv}$ for the PC-based device, while light-colored traces represent $S_{vv}$ for the mirror-based device. The low voltage noise background $S_{uu}$ observed for the PC-based device results from the low reflectance $R$, as shown in Fig.~2e and described by Eq.~(\ref{Suu}). At $V_\mathrm{gate}=0$, the thermal resonance of the mirror-based device at $\lambda=637$~nm is barely visible, while that of the PC-based device is clearly resolved, with $\mathrm{SNR}\simeq17$~dB. At $V_\mathrm{gate}=5$~V, the increase in $f_0$ means that SNR is reduced. To compensate, we choose to measure at $\lambda=672$~nm, where $\mathcal{R}$ is higher at lower $z_\mathrm{m}$.  Under these conditions, we find that the thermal resonance cannot be resolved in the mirror-based device, while it is still clearly visible in the PC-based device, with $\mathrm{SNR}\simeq10$~dB.

The temperature increase $\Delta T$ due to absorptive heating is estimated by solving the heat equation and averaging the resulting temperature over the resonator \cite{Barton2012,Storch2018}. This estimation assumes that the light is focused into a spot much smaller than the diameter of the resonator. It reads:
\begin{equation}
\Delta T\simeq\frac{ A\times P_\mathrm{inc}}{4\pi\kappa d_\mathrm{G}}\,,\label{DeltaT}
\end{equation}
with $\kappa$ the in-plane thermal conductivity of graphene and $d_\mathrm{G}$ the thickness of graphene. At room temperature, the values of $\kappa\simeq1000$--$5000$~W~m$^{-1}$~K$^{-1}$ reported in Refs.~\cite{Balandin2008,Ghosh2010,Cai2010,Faugeras2010,JULee2011,ShanshanChen2012,Pop2012,XXu2014,QYLi2017} result in $\Delta T\simeq0.3$--$0.06$~K, indicating that heating is negligible. Furthermore, we have verified that the coupling between vibrations and photothermal forces also causes a negligible change in temperature.

By combining $R$, $\mathcal{R}$, $A$, $\partial A/\partial z_\mathrm{m}$, $f_0$, and $z_\mathrm{m}$, we calculate SNR as a function of $V_\mathrm{gate}$ for both $\lambda=637$~nm and $\lambda=672$~nm. The result, shown in Fig.~3d, indicates that the ability to choose between these two wavelengths ensures that the SNR remains above 9~dB across a broad range of $z_\mathrm{m}$. In turn, the clear observation of thermal vibrations means that the device, employed as a force detector, has reached its sensitivity limit set by Langevin forces. This limit is $S^{1/2}(V_\mathrm{gate})=[8\pi k_\mathrm{B}Tmf_0(V_\mathrm{gate})/Q]^{1/2}\simeq(4.3$--$5.5)\times10^{-16}$~N~Hz$^{-1/2}$, for $V_\mathrm{gate}$ varying from 0 to 5~V.

Implementing the PC-based reflectometry technique at cryogenic temperatures may require lowering $P_\mathrm{inc}$ to mitigate temperature changes caused by absorptive heating and optomechanical effects \cite{Metzger2004,Zaitsev2011,Barton2012}. Heating is a primary concern owing to the large absorption of optical energy and the reduced in-plane thermal conductivity of graphene at low temperatures. At 10~K, $\kappa$ may be as high as 100~W~m$^{-1}$~K$^{-1}$, as in graphite \cite{Smith1954,Morelli1985,Nakamura2017}, but it may also be much lower \cite{XXu2014,XWei2014}. For example, assuming $\kappa=10$~W~m$^{-1}$~K$^{-1}$ and $P_\mathrm{inc}=0.1\times10^{-6}$~W results in $\Delta T\simeq1.5$~K which, if unaccounted for, could lead to a 15\% overestimation of the force sensitivity $S$. Assuming a quality factor $Q=10^4$, the thermal resonance would still be observable, with $\mathrm{SNR}\simeq5$~dB. However, any further reduction in $\kappa$ would impair force sensing measurements, at least based on the model used to derive Eq.~(\ref{DeltaT}). While thermal vibrations of 2D resonators have been measured at cryogenic temperatures in some instances (see Table~I in Appendix), we believe that heating of vibrational modes remains a practical concern.

\section{Conclusion and outlook}

Our simulations demonstrate that a photonic crystal (PC) structure enables sensitive optical reflectometry of the thermal vibrations of a graphene resonator, even under low incident light intensities where a perfectly reflective mirror is ineffective. Our PC-based detection method achieves a maximum responsivity that is 10 times higher than the mirror-based approach. Furthermore, the two guided resonances supported by the PC provide a high signal-to-noise ratio across a broad range of static membrane displacements. Looking ahead to experimental implementation, future work will investigate how small variations in hole size within the PC lattice affect the responsivity. Quantifying the variance of vibrational amplitude fluctuations as a function of bath temperature will help define the limits of the technique's applicability at cryogenic temperatures.

\begin{acknowledgments}
This work was supported by the National Natural Science Foundation of China (grant numbers 62150710547 and 62074107) and the project of the Priority Academic Program Development (PAPD) of Jiangsu Higher Education Institutions.
\end{acknowledgments}

\section*{Data availability}
The data that support the findings of this article will be made openly available.

\section*{Appendix}

Table~I presents a survey of experimental studies in which the thermal vibrations of two-dimensional nanomechanical resonators have been measured. The table is organized as follows, from left to right: full reference; measurement temperature (RT stands for room temperature); measurement technique; power of the probe incident on the resonator (entries marked with --- indicate the value has not been reported); resonator material; thickness of the resonator; and the resonant frequencies of the measured vibrational modes.
\begin{table}[htbp]
    \scriptsize
    \centering 
	\caption{A survey of experimental studies in which the thermal vibrations of two-dimensional nanomechanical resonators have been measured. RT: room temperature. Opt. reflect.: optical reflectometry. Microwave opt.: microwave optomechanics. Michelson interf.: Michelson interferometry. Cavity opt.: optical cavity optomechanics. Frequencies: resonant frequencies of the thermal vibrations, including the fundamental mode and higher order modes.}
	\label{tab:shape-functions}
	\begin{tabular}{ccccccccc}
		\hline
		Ref. & Temperature & Readout & Power incident on the resonator& Material & Thickness & Frequencies (MHz)\\
		\hline
		\cite{Bunch2007} & RT & Opt. reflect. & --- & Graphene & 5~nm & 36\\
		\cite{Barton2012} & RT & Opt. reflect. & 1~mW & Graphene & Monolayer & 5\\
		\cite{Lee2013} & RT & Opt. reflect. & 100--700~$\mu$W & MoS$_2$ & 6--62~nm & 48--58\\
		\cite{RvanLeeuwen2014} & RT & Opt. reflect. & 0.2--3~$\mu$W & MoS$_2$ & 5 layers & 7.5\\
		\cite{Wang2014} & RT & Opt. reflect. & $<700$~$\mu$W & MoS$_2$ & 60~nm & 14--54\\
		\cite{Wang2015} & RT & Opt. reflect. &700~$\mu$W & Black Phosphorus & 22--190~nm & 5--11\\
		\cite{Lee2016} & RT & Opt. reflect. & --- & MoS$_2$ & 31--81~nm & 16--22\\
		\cite{Davidovikj2016} & RT & Opt. reflect. &800~$\mu$W & Graphene & 5~nm & 13--23\\
		\cite{Zheng2017hexagonal} & RT & Opt. reflect. & --- & hexagonal Boron Nitride & 6.7-- 292~nm& 5--70\\
	       \cite{Zheng2017ultrawide} & RT & Opt. reflect. & 2.8~mW & $\beta$-Ga$_2$O$_3$ & 20--140~nm & 10--75\\
		\cite{Lee2018} & RT & Opt. reflect. & 350~$\mu$W & MoS$_2$ & 1--4~layers & 28--43\\
	   	\cite{Islam2018} & RT & Opt. reflect. & 1.6~mW & Black Phosphorus & 10--100~nm & 7--22\\	   
		\cite{Kim2018} & RT & Opt. reflect. & 50~$\mu$W & Graphene+MoS$_2$ & Monolayer & 11.6\\
		\cite{Zheng2018} & RT & Opt. reflect. & --- & $\beta$-Ga$_2$O$_3$ & 73~nm & 30\\
		\cite{Dolleman2019} & 345~K & Opt. reflect. & 2~mW (before the objective) & Graphene & Monolayer & 14\\
		\cite{ZhangXin2020} & RT & Opt. reflect. & 500~$\mu$W & Graphene & Monolayer & 33\\
		\cite{Wang2021} & RT & Opt. reflect. & 350~$\mu$W & MoS$_2$ & 56~nm & 22\\
		\cite{Lee2021} & RT & Opt. reflect. & 350~$\mu$W & MoS$_2$ & 31~nm & 20\\
		\cite{Zhu2022} & RT & Opt. reflect. & --- & WSe$_2$ & 18.4~nm & 3.3\\
		\cite{Yang2022} & RT & Opt. reflect. & --- & MoS$_2$ & Monolayer & 35.3\\
		\cite{Kaisar2022} & RT & Opt. reflect. & 350~$\mu$W & MoS$_2$ & Monolayer & 29\\
		\cite{Lee2022} & RT & Opt. reflect. & 700~$\mu$W & MoS$_2$ & 1--13 layers & 33--60\\
		\cite{Zhang2024} & RT & Opt. reflect. & --- & MoS$_2$ & 2 layers &19\\
		\cite{Yousuf2024} & RT & Opt. reflect. & --- & MoS$_2$ & 2 layers & 18--21\\
	    	\cite{Lee2025} & RT & Opt. reflect. & 40--125~$\mu$W & Graphene & 2 layers & 1.3, 19\\
	   	\cite{Yang2025} & RT & Opt. reflect. & 5~$\mu$W & Graphene & 10 layers & 16\\
		\cite{Morell2016} & 20~K & Opt. reflect. & 10~$\mu$W & WSe$_2$ & monolayer & 58\\
		\cite{DeAlba2016} & RT & Opt. reflect. & 130~$\mu$W  & Graphene & monolayer & 2.95\\
		\cite{JWFang2025} & 5~K & Opt. reflect. & 150~$\mu$W  & CrPS$_4$ & 25~nm & 77\\
		\cite{Schwarz2016} & RT & Michelson interf. & 1--100~$\mu$W  & Graphene & monolayer & 2.3\\
		\cite{Singh2018} & RT & Michelson interf. & 100~$\mu$W & Graphene & Monolayer & 4.15\\
		\cite{Singh2020} & RT & Michelson interf. & 400~$\mu$W & Graphene & Monolayer & 2.75\\
		\cite{Cole2015} & RT & Cavity opt. & 1.47~$\mu$W (absorbed)  & Graphene & monolayer & 1.4--4\\
	      \cite{Singh2014} & 96~mK & Microwave opt. & $7\times10^5$~photons & Graphene & 10~nm & 36\\
	      \cite{Weber2016} & 40~mK & Microwave opt. & $6\times10^4$~photons & Graphene & 25 layers & 53.7\\
		\cite{Guettinger2017} &15~mK & Microwave opt. & $1.9\times10^5$~photons & Graphene & 6 layers& 46\\
	      \cite{Kumar2015} & RT & Piezoresistance & --- & Graphene & Monolayer & 1.7\\
		\hline
	\end{tabular}
\end{table}

\newpage

\end{document}